\documentstyle[twoside,fleqn,espcrc2,epsf,epsfig]{article}

\newcommand{\ewxy}[2]{\setlength{\epsfxsize}{#2}\epsfbox[10 60 640 570]{#1}}

\title{The spectrum from Lattice NRQCD}
\author{Presented by C.~T.~H.~Davies\address{Dept. of Physics and Astronomy,
University of Glasgow, Glasgow, G12 8QQ, Scotland}
\thanks{Work supported by PPARC, NATO under grant 941259
and EU network grant CHRX-CT92-0051}}

\begin{document}

\begin{abstract}
I review recent results for heavy-heavy spectroscopy using Lattice NRQCD.
The NRQCD collaboration reports that spin-independent splittings for
the $\Upsilon$ are
scaling for a sensible range of $\beta$ values in the quenched
approximation. Spin-dependent
splittings are not, if the scale is set by spin-independent
splittings. Results which include
higher order spin-dependent relativistic and discretisation
corrections show differences from previous (NRQCD collaboration)
results without these.
As expected, the differences are small for $\Upsilon$ but rather large for
charmonium. New results from the SESAM collaboration for
 $\Upsilon$ spectroscopy on
 configurations with Wilson
dynamical fermions show good agreement with previous results on
HEMCGC configurations with staggered dynamical fermions.
\end{abstract}

\maketitle
\section{New results}

The new results using NRQCD which I will discuss are :
\begin{itemize}
\item SESAM collaboration results for the $\Upsilon$
spectrum on configurations with 2 flavours of dynamical Wilson quarks.
\item Results from the NRQCD collaboration for the $\Upsilon$ spectrum
on UKQCD quenched configurations
at $\beta$ = 6.2, giving a more detailed spectrum
than previous results at this fine lattice spacing.
\item Results for the $\Psi$ and $\Upsilon$ spectrum which include additional
relativistic and discretisation corrections in spin-dependent terms
from Trottier, SESAM and UKQCD.
\item Further $B_c$ results on dynamical configurations and including
relativistic $c$ quarks with non-relativistic $b$s.
\end{itemize}

\section{NRQCD}

The splittings between radial and orbital excitations for systems made
of heavy quarks are around 500 MeV, much less than the masses of the bound
states. This implies that these are non-relativistic systems and a
systematic expansion of the QCD Hamiltonian in powers of $v^2$ may be
useful ~\cite{bat}.

The continuum action density, correct
through $\cal{O}$$(M_Qv^4)$, is broken down according to

\begin{equation}
{\cal L}_{cont}=\psi^{\dagger}(D_t+H_0^{cont})\psi+\psi^{\dagger}
\delta H^{cont}\psi \label{deltaH}
\end{equation}

\noindent
$H_0^{cont}$ and $\delta H^{cont}$ are given explicitly in ref.\cite{nak}.
On the lattice $H_0$ and the leading piece of $\delta H$, $\delta H_1$
are given by:

$$
H_0=-\frac{\Delta^{(2)}}{2M_Q^0} \qquad\mbox{and}$$

\begin{eqnarray*}
\delta H_1 &\hspace{-0.5em} = &\hspace{-1em}{} -c_1\frac{(\Delta^{(2)})^2}
{8(M_Q^0)^3}+c_2
\frac{ig}{8(M_Q^0)^2}
({\bf \Delta\cdot E}-{\bf E\cdot\Delta})\\ &&\hspace{-1em}{} -c_3
\frac{g}{8(M_Q^0)^2}{\bf \sigma\cdot
(\Delta}\times{\bf E}-{\bf E}\times{\bf \Delta})\\ &&\hspace{-1em}{} -c_4
\frac{g}{2M_Q^0}{\bf \sigma
\cdot B}+c_5\frac{a^2\Delta^{(4)}}{24M_Q^0}-c_6
\frac{a(\Delta^{(2)})^2}{16n(M_Q^0)^2} \label{lattice1}
\end{eqnarray*}

\noindent
The last two terms in $\delta H_1$ come from finite lattice spacing corrections
to the lattice Laplacian and lattice time derivative, of
$\cal{O}$$(a^2M_Q^3v^4)$
and $\cal{O}$$(aM_Q^2v^4)$ respectively. $n$ is
the stability parameter used in the evolution equation below.

The quark propagators are determined from evolution equations that specify the
propagator value, for $t>0$, in terms of the value on the previous timeslice;

\begin{eqnarray*}
G_1&\hspace{-1em}= &\hspace{-1em}{\left(1-\frac{aH_0}{2n}\right)}^n
U_4^{\dagger}
     {\left(1-\frac{aH_0}{2n}\right)}^n \delta_{x,0}\mbox{,}\\
G_{t+1}&\hspace{-1em}= &\hspace{-1em}{\left(1-\frac{aH_0}{2n}\right)}^n
U_4^{\dagger}
     {\left(1-\frac{aH_0}{2n}\right)}^n(1-a\delta H)G_t \label{evolve}
\end{eqnarray*}

The quark propagators are combined with smearing operators at source
and sink to produce good overlap with different states.
This is done in different ways by different groups, NRQCD and SESAM
using Coulomb gauge wavefunction smearing and UKQCD, gauge-invariant
blocking.
All use multi-exponential fits to multiple correlation functions on
large ensembles to obtain masses for radial excitations in $s$ and $p$
channels which can be compared to experiment.

\begin{figure}[h]
\begin{center}
\setlength{\unitlength}{.02in}
\begin{picture}(130,220)(10,850)
\put(15,935){\line(0,1){125}}
\multiput(13,950)(0,50){3}{\line(1,0){4}}
\multiput(14,950)(0,10){10}{\line(1,0){2}}
\put(12,950){\makebox(0,0)[r]{9.5}}
\put(12,1000){\makebox(0,0)[r]{10.0}}
\put(12,1050){\makebox(0,0)[r]{10.5}}
\put(12,1070){\makebox(0,0)[r]{GeV}}
\put(15,935){\line(1,0){115}}


\put(27,930){\makebox(0,0)[t]{${^1S}_0$}}

\put(25,943.1){\circle{2}}
\put(25,1002){\circle{2}}
\put(30,942){\circle*{2}}

\put(33,942.7){\makebox(0,0){$\star$}}
\put(33,942.7){\line(0,1){0.1}}
\put(33,942.7){\line(0,-1){0.1}}
\put(33,999.1){\makebox(0,0){$\star$}}
\put(33,999.1){\line(0,1){2.5}}
\put(33,999.1){\line(0,-1){1.9}}
\put(33,1055.1){\makebox(0,0){$\star$}}
\put(33,1055.1){\line(0,1){12.5}}
\put(33,1055.1){\line(0,-1){11.2}}

\put(35,942.6){\makebox(0,0){$\ast$}}
\put(35,942.6){\line(0,1){0.2}}
\put(35,942.6){\line(0,-1){0.1}}
\put(35,1002.6){\makebox(0,0){$\ast$}}
\put(35,1002.6){\line(0,1){2.4}}
\put(35,1002.6){\line(0,-1){2.8}}
\put(35,1045.7){\makebox(0,0){$\ast$}}
\put(35,1045.7){\line(0,1){11.1}}
\put(35,1045.7){\line(0,-1){10.6}}


\put(58,930){\makebox(0,0)[t]{${^3S}_1$}}
\put(75,946){\makebox(0,0){{\small 1S}}}
\multiput(43,946)(3,0){9}{\line(1,0){2}}
\put(75,1002){\makebox(0,0){{\small 2S}}}
\multiput(43,1002)(3,0){9}{\line(1,0){2}}
\put(75,1036){\makebox(0,0){{\small 3S}}}
\multiput(43,1036)(3,0){9}{\line(1,0){2}}

\put(50,946){\circle{2}}
\put(50,1004.1){\circle{2}}
\put(50,1060){\circle{2}}
\put(50,1060){\line(0,1){4.8}}
\put(50,1060){\line(0,-1){4.8}}

\put(55,946){\circle*{2}}
\put(55,1003){\circle*{2}}
\put(55,1004){\line(0,1){1.4}}
\put(55,1002){\line(0,-1){1.4}}
\put(55,1039.1){\circle*{2}}
\put(55,1039.1){\line(0,1){7.2}}
\put(55,1039.1){\line(0,-1){7.2}}

\put(60,946){\makebox(0,0){$\star$}}
\put(60,1001.3){\makebox(0,0){$\star$}}
\put(60,1001.3){\line(0,1){2.1}}
\put(60,1001.3){\line(0,-1){1.9}}
\put(60,1048.0){\makebox(0,0){$\star$}}
\put(60,1048.0){\line(0,1){10.2}}
\put(60,1048.0){\line(0,-1){9.1}}

\put(63,946){\makebox(0,0){$\ast$}}
\put(63,1002.6){\makebox(0,0){$\ast$}}
\put(63,1002.6){\line(0,1){1.9}}
\put(63,1002.6){\line(0,-1){1.8}}
\put(63,1041.8){\makebox(0,0){$\ast$}}
\put(63,1041.8){\line(0,1){9.6}}
\put(63,1041.8){\line(0,-1){9.1}}


\put(105,930){\makebox(0,0)[t]{$^1P_1$}}

\put(115,990){\makebox(0,0){{\small 1P}}}
\multiput(83,990)(3,0){9}{\line(1,0){2}}
\put(115,1026){\makebox(0,0){{\small 2P}}}
\multiput(83,1026)(3,0){9}{\line(1,0){2}}

\put(90,987.6){\circle{2}}
\put(90,1038.7){\circle{2}}
\put(90,1038.7){\line(0,1){4.8}}
\put(90,1038.7){\line(0,-1){4.8}}

\put(95,989){\circle*{2}}
\put(95,1023){\circle*{2}}
\put(95,1023){\line(0,1){7.2}}
\put(95,1023){\line(0,-1){7.2}}

\put(100, 988.7){\makebox(0,0){$\star$}}
\put(100, 988.7){\line(0,1){2.3}}
\put(100, 988.7){\line(0,-1){2.4}}
\put(100,1041.7){\makebox(0,0){$\star$}}
\put(100,1041.7){\line(0,1){7.4}}
\put(100,1041.7){\line(0,-1){8.6}}

\put(103, 989.8){\makebox(0,0){$\ast$}}
\put(103, 989.8){\line(0,1){2.4}}
\put(103, 989.8){\line(0,-1){1.4}}
\put(103,1030.5){\makebox(0,0){$\ast$}}
\put(103,1030.5){\line(0,1){6.7}}
\put(103,1030.5){\line(0,-1){6.2}}

\multiput(0,900)(3,0){3}{\line(1,0){2}}
\put(10,900){\makebox(0,0)[l]{: {Experiment}}}
\put(2,890){\circle{2}}
\put(10,890){\makebox(0,0)[l]{: {NRQCD $(n_f = 0) , \beta =
      6.0 $}}}
\put(2,880){\circle*{2}}
\put(10,880){\makebox(0,0)[l]{: {NRQCD $(n_f = 2, KS, am_q = 0.01) ,
        \beta = 5.6 $}}}
\put(0,870){\makebox(0,0)[tl]{$\star$}}
\put(10,870){\makebox(0,0)[l]{: {SESAM $(n_f = 2, W, \kappa = 0.157),
        \beta = 5.6 $}}}
\put(0,860){\makebox(0,0)[tl]{$\ast$}}
\put(10,860){\makebox(0,0)[l]{: {SESAM $(n_f = 2, W, \kappa = 0.1575),
        \beta = 5.6 $}}}

\end{picture}
\hspace{1.5cm}
\end{center}
\caption{The spin independent spectrum for $b\overline{b}$}
\end{figure}

\begin{eqnarray*}
\delta H_2 &\hspace{-0.5em} = &\hspace{-1em}{} - f_1\frac {g}
{8(M_Q^0)^3} \{ \Delta^{(2)} , \sigma \cdot {\bf B} \} \\
&&\hspace{-1em}{} - f_2  \frac {3g}
{64(M_Q^0)^4} \{ \Delta^{(2)} , \sigma \cdot
({\bf \Delta}\times{\bf E}-{\bf E}\times{\bf \Delta})\} \\ &&\hspace{-1em}{}
- f_3 \frac {ig^2} {8(M_Q^0)^3} \sigma \cdot {\bf E \times E}  \label{deltaH2}
\end{eqnarray*}

to be added to $\delta H_1$ above in the evolution equation.
The action used is then correct through $\cal{O}$$(M_Qv^6)$ for
{\it spin-dependent} terms. In addition these groups
have included extra discretisation corrections for the spin-dependent
terms in $\delta H_1$ at $\cal{O}$$(M_Qv^4)$, i.e. terms of
$\cal{O}$$(a^2M_Q^3v^6)$.
 These involve ~\cite{nak} replacing the $E$ and
$B$ fields with an improved version:
\begin{eqnarray*}
\tilde{F}_{\mu \nu} &=& \frac {5} {3} F_{\mu \nu} -
\frac {1} {6} [ U_{\mu}(x) F_{\mu \nu} (x+\hat{\mu}) U^{\dagger}_{\mu}(x)\\
&+& U^{\dagger}_{\mu}(x-\hat{\mu}) F_{\mu \nu} (x-\hat{\mu})
U_{\mu}(x-\hat{\mu}) \\
&+& ( \mu \leftrightarrow \nu) ]
\end{eqnarray*}
 and replacing the spatial
derivative with an improved version (as was done for the leading
spin-independent terms
in $\delta H_1$) :
\begin{equation}
\tilde{\Delta}^{(2)} = \Delta^{(2)} - \frac {a^2} {12} \Delta^{(4)}.
\end{equation}
These corrections then appear as additional terms in $\delta H_2$.
The results discussed here differ in how groups
 have treated $E$ and $B$ fields and
corrections to derivatives in the spin-independent terms at
$\cal{O}$$(M_Qv^4)$. This should not cause a big effect since
these corrections are next-to-next-to-leading order in
spin-independent splittings.

\begin{figure}[h]
\begin{center}
\setlength{\unitlength}{.02in}
\begin{picture}(100,200)(15,-150)
\put(15,-50){\line(0,1){80}}
\multiput(13,-40)(0,20){4}{\line(1,0){4}}
\multiput(14,-40)(0,10){7}{\line(1,0){2}}
\put(12,-40){\makebox(0,0)[r]{$-40$}}
\put(12,-20){\makebox(0,0)[r]{$-20$}}
\put(12,0){\makebox(0,0)[r]{$0$}}
\put(12,20){\makebox(0,0)[r]{$20$}}
\put(12,40){\makebox(0,0)[r]{MeV}}
\put(15,-50){\line(1,0){110}}

\multiput(28,0)(3,0){9}{\line(1,0){2}}
\put(60,2){\makebox(0,0)[t]{$\Upsilon$}}
\put(60,-34){\makebox(0,0)[t]{$\eta_b$}}

\put(35,0){\circle{2}}
\put(35,-29.9){\circle{2}}

\put(30,0){\makebox(0,0){$\bowtie$}}
\put(30,-25.7){\makebox(0,0){$\bowtie$}}

\put(40,0){\circle*{2}}
\put(40,-39){\circle*{2}}
\put(40,-39){\line(0,1){2}}
\put(40,-39){\line(0,-1){2}}

\put(45,0){\makebox(0,0){$\star$}}
\put(45, -33.2){\makebox(0,0){$\star$}}
\put(45, -33.2){\line(0,1){1.5}}
\put(45, -33.2){\line(0,-1){1.9}}

\put(48,0){\makebox(0,0){$\ast$}}
\put(48, -33.7){\makebox(0,0){$\ast$}}
\put(48, -33.7){\line(0,1){1.6}}
\put(48, -33.7){\line(0,-1){1.5}}

\put(63,-10){\makebox(0,0)[l]{$h_b$}}
\put(70,-1.8){\circle{2}}
\put(75,-2.9){\circle*{2}}
\put(75,-2.9){\line(0,1){1.2}}
\put(75,-2.9){\line(0,-1){1.2}}


\put(80,  -1.0){\makebox(0,0){$\star$}}
\put(80,  -1.0){\line(0,1){1.5}}
\put(80,  -1.0){\line(0,-1){2.2}}
\put(85,   0.3){\makebox(0,0){$\ast$}}
\put(85,   0.3){\line(0,1){1.2}}
\put(85,   0.3){\line(0,-1){1.5}}

\multiput(90,-40)(3,0){9}{\line(1,0){2}}
\put(118,-40){\makebox(0,0)[l]{$\chi_{b0}$}}
\multiput(90,-8)(3,0){9}{\line(1,0){2}}
\put(118,-8){\makebox(0,0)[l]{$\chi_{b1}$}}
\multiput(90,13)(3,0){9}{\line(1,0){2}}
\put(118,13){\makebox(0,0)[l]{$\chi_{b2}$}}

\put(97,-25.1){\circle{2}}
\put(97,-8.6){\circle{2}}
\put(97,10.2){\circle{2}}

\put(102,-34){\circle*{2}}
\put(102,-34){\line(0,1){5}}
\put(102,-34){\line(0,-1){5}}
\put(102,-7.9){\circle*{2}}
\put(102,-7.9){\line(0,1){2.4}}
\put(102,-7.9){\line(0,-1){2.4}}
\put(102,11.5){\circle*{2}}
\put(102,11.5){\line(0,1){2.4}}
\put(102,11.5){\line(0,-1){2.4}}

\put(92,-19){\makebox(0,0){$\bowtie$}}
\put(92,-19){\line(0,1){3}}
\put(92,-19){\line(0,-1){3}}
\put(92,-3.8){\makebox(0,0){$\bowtie$}}
\put(92,-3.8){\line(0,1){2}}
\put(92,-3.8){\line(0,-1){2}}
\put(92,6.5){\makebox(0,0){$\bowtie$}}
\put(92,6.5){\line(0,1){1.3}}
\put(92,6.5){\line(0,-1){1.3}}

\put(106, -36.9){\makebox(0,0){$\star$}}
\put(106, -36.9){\line(0,1){3.5}}
\put(106, -36.9){\line(0,-1){4.2}}
\put(106, -10.1){\makebox(0,0){$\star$}}
\put(106, -10.1){\line(0,1){2.9}}
\put(106, -10.1){\line(0,-1){3.7}}
\put(106,   7.8){\makebox(0,0){$\star$}}
\put(106,   7.8){\line(0,1){4.7}}
\put(106,   7.8){\line(0,-1){4.8}}

\put(110, -32.3){\makebox(0,0){$\ast$}}
\put(110, -32.3){\line(0,1){3.6}}
\put(110, -32.3){\line(0,-1){3.8}}
\put(110,  -6.3){\makebox(0,0){$\ast$}}
\put(110,  -6.3){\line(0,1){1.5}}
\put(110,  -6.3){\line(0,-1){1.7}}
\put(110,  10.2){\makebox(0,0){$\ast$}}
\put(110,  10.2){\line(0,1){1.9}}
\put(110,  10.2){\line(0,-1){1.5}}

\multiput(0,-60)(3,0){3}{\line(1,0){2}}
\put(10,-60){\makebox(0,0)[l]{: {Experiment}}}
\put(0,-68){\makebox(0,0)[tl]{$\bowtie$}}
\put(10,-70){\makebox(0,0)[l]{: {Manke {\it et al} (UKQCD)  $(n_f = 0),$}}}
\put(60,-80){\makebox(0,0)[l]{ $\beta = 6.0 , \delta H_1 + \delta H_2 ,
u_{0P}$}}
\put(2,-90){\circle{2}}
\put(10,-90){\makebox(0,0)[l]{: {NRQCD $(n_f = 0) ,$}}}
\put(60, -100){\makebox(0,0)[l]{ $ \beta =
      6.0, \delta H_1, u_{0P}$ }}
\put(2,-110){\circle*{2}}
\put(10,-110){\makebox(0,0)[l]{: {NRQCD $(n_f = 2, KS, am_q=0.01) ,$}}}
\put(60,-120){\makebox(0,0)[l]{ $ \beta = 5.6 , \delta H_1 , u_{0P}$}}
\put(0,-130){\makebox(0,0)[tl]{$\star$}}
\put(10,-130){\makebox(0,0)[l]{: {SESAM $(n_f = 2, W, \kappa = 0.157),$}}}
\put(60,-140){\makebox(0,0)[l]{ $ \beta = 5.6 , \delta H_1 + \delta H_2 ,
u_{0L}$}}
\put(0,-150){\makebox(0,0)[tl]{$\ast$}}
\put(10,-150){\makebox(0,0)[l]{: {SESAM $(n_f = 2, W, \kappa = 0.1575),$}}}
\put(60,-160){\makebox(0,0)[l]{ $ \beta = 5.6 , \delta H_1 + \delta H_2 ,
u_{0L}$}}

\end{picture}
\end{center}
\caption{$\Upsilon$ fine structure}
\end{figure}
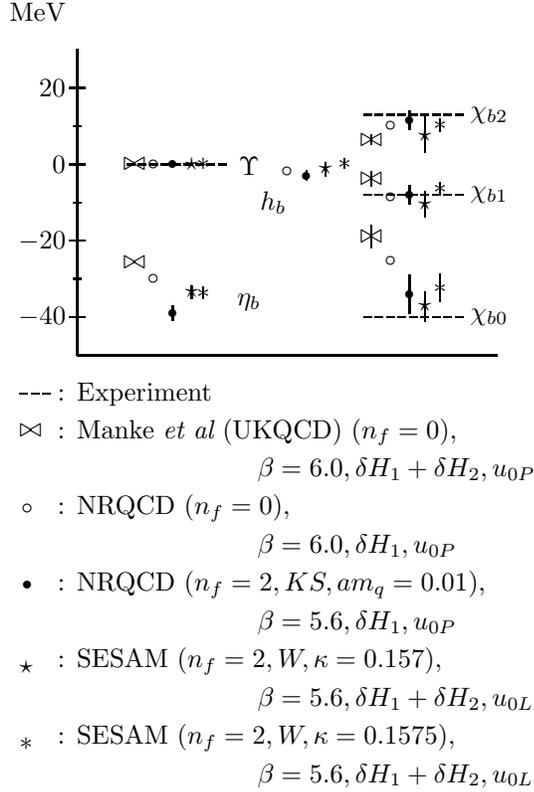

The $c_i$ and $f_i$ coefficients should be matched to full QCD
either perturbatively or non-perturbatively. This has not been
done by any of the groups. Instead they have relied upon tadpole-dominance
arguments ~\cite{landm} to replace the gauge fields $U_{\mu}$ appearing in
derivatives
and $E$ and $B$ fields by $U_{\mu}/u_0$. $u_0$ represents the effect
of tadpoles in reducing the mean value of the link. Different values
can be taken for $u_0$ and they will be compared below. After
these modified gauge links are used in the action the $c_i$ and
$f_i$ are set their tree-level values of 1. There is
evidence from perturbative calculations of the NRQCD self-energy diagram
that the radiative corrections to this, at least for $c_1$ and $c_5$, are
very small (less than 10\%) for reasonable values of
$M_Qa$ ~\cite{morningstar}. However, they could in principle
be of the
same size as the relativistic and discretisation corrections of
$\delta H_2$.

There are two parameters then to fix in the lattice calculation:
$a$ and $M_Q$. For $a$, the 2S-1S or 1P-1S splitting should be used,
where S and P are spin averages over $s$ and $p$ states.
This has the advantage of being independent of quark mass experimentally
in the $b$, $c$ region. Because not all the states have been seen
experimentally the $2{^3S_1} - 1{^3S_1}$ and
$1^3\overline{P} - 1{^3S_1}$ splittings are used in the $b\overline{b}$
sector, where the $^3\overline{P}$ is the spin average
of the $^3P_{0,1,2}$.  To fix the quark masses the energy
at finite momentum
is calculated for one meson (say the $1 ^3S_1$) and the denominator of the
kinetic term in the
dispersion relation is taken as the absolute mass of that
meson in lattice units. The energy at zero momentum differs
from this since the mass term was dropped from $H_0$. The difference between
the energy at zero momentum and the mass can be compared to
perturbative predictions and agrees well ~\cite{ups}.

\section{Results for Upsilon spectroscopy}

The results to be discussed are tabulated below.
Where a symbol is denoted in the last column that
is the one used in Figures 1 and 2.

\begin{table}
\begin{tabular}{llllll}
$\beta, n_f$ & V.T & configs: & results: & Fig. \\
\hline \\
5.7,0 & $12^3.24$ & UKQCD & NRQCD & \\
6.0,0 & $16^3.32$ & Kogut {\it et al} & NRQCD & $\circ$ \\
6.2,0 & $24^3.32$ & UKQCD & NRQCD & \\
6.0,0 & $16^3.48$ & UKQCD & UKQCD & $\bowtie$ \\
5.6,2 & $16^3.32$ & HEMCGC & NRQCD & $\bullet$ \\
5.6,2 & $16^3.32$ & SESAM & SESAM & $\star$,$\ast$ \\
&&&& \\
\end{tabular}
\caption{ Parameters for the ensembles used in the results discussed.}
\end{table}

The HEMCGC configurations use Kogut-Susskind dynamical
fermions and have two ensembles, one with $ma$ = 0.01
and one with $ma$ = 0.025. The NRQCD results on these
configurations have been previously reported ~\cite{alpha}.
The matching quenched results at $\beta$ = 6.0 represent
a higher statistics study than that reported in ref.~\cite{ups}.
The SESAM configurations use
Wilson dynamical fermions and have three ensembles. The two
with lightest dynamical mass ($\kappa$ = 0.157 and 0.1575) were
used for the $\Upsilon$ spectrum calculations by Achim Spitz and
Henning Hoeber.
SESAM results on these configurations
are reported elsewhere in this Proceedings ~\cite{lippert}.

\begin{figure}
\centerline{\ewxy{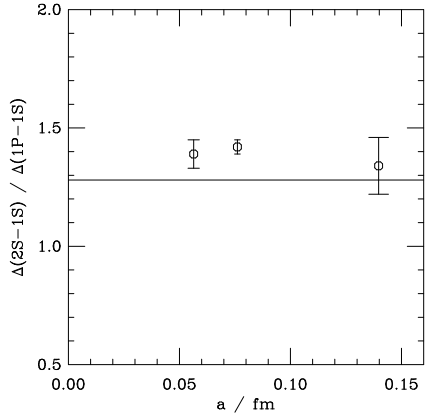}{75mm}}
\caption{
The ratio of splittings $2^3S_1 - 1^3S_1$ to $1^3\overline{P} -
1^3S_1$ in $b\overline{b}$ as a function of lattice spacing in fm at 3
different
values of the lattice spacing on quenched configurations
(NRQCD collaboration). The solid
line represents the experimental result.
}
\end{figure}

Figure 1 shows the collected results for the
spin-independent spectrum
and Figure 2 the fine structure.
The values of the lattice spacing used to convert to
physical units have been taken as the average of that
from the 2S-1S and 1P-1S splittings. $a^{-1}$ takes the
value  2.4 GeV in all cases except for the heavier dynamical mass
SESAM results in which it is 2.3 GeV. A bare $b$ quark
mass in lattice units of 1.71 was obtained by the NRQCD
collaboration by tuning
on the quenched $\beta$ = 6.0 configurations. A value of 1.8
was used by them on the HEMCGC unquenched configurations but
results from a kinetic mass analysis of the $\Upsilon$ ~\cite{nfdep}
indicate that this was too large and 1.7 would have been better.
The SESAM results use a bare $b$ mass of 1.7.

{\bf Spin-independent spectrum} - It is clear from the open circles
in Figure 1 that the spin-independent spectrum on quenched configurations
is not correct. We expect there to be errors because the coupling
constant runs incorrectly between the scales appropriate to, say,
the 1P and the 2S, so that it is not possible to fix an effective coupling
which gives the correct answer for both states. Before comparing
results at different values of $n_f$, however, we must first check
that the results are scaling for a given value of $n_f$.

\begin{figure}
\centerline{\ewxy{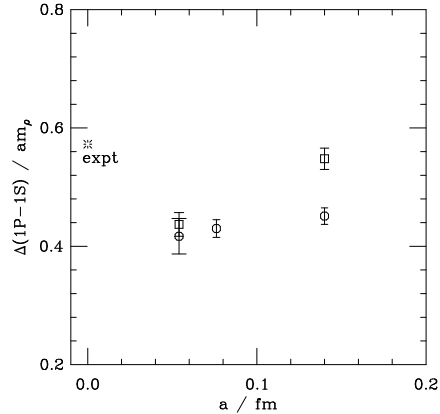}{75mm}}
\vspace{0.2in}
\caption{ The ratio of the $1^3\overline{P} - 1^3S_1$ splitting
in $b\overline{b}$ to the $\rho$ mass as a function of lattice spacing in fm at
3 different values of the lattice spacing in the quenched
approzimation. $\Upsilon$
results from the NRQCD collaboration. The circles use
UKQCD results for the $\rho$ mass, the squares GF11 results.}
\end{figure}

NRQCD is an effective theory reproducing the low energy behaviour of
QCD but in which the ultra-violet cut-off plays a crucial r\^ole. It is
therefore not possible to take $a$ to zero within NRQCD. The $c_i$ and
$f_i$ coefficients will start to diverge as powers of $1/M_Qa$
{}~\cite{morningstar} and
we will lose control of the NRQCD expansion. However, there is no
need to take $a$ to zero if we can demonstrate $a$ independence of our
results for a reasonable range of values of $a$. Provided $M_Qa > 0.6$
the $c_i$ coefficients which have been calculated are perfectly well
behaved ~\cite{morningstar} and we do not expect any problems.
 For the $\Upsilon$ system
this corresponds to $\beta < 6.4$.

\begin{figure}
\centerline{\ewxy{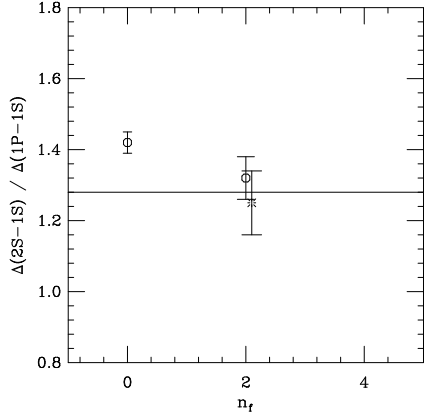}{75mm}}
\vspace{0.2in}
\caption{
The ratio of splittings $2^3S_1 - 1^3S_1$ to $1^3\overline{P} -
1^3S_1$ in $b\overline{b}$ as a function of the number of
dynamical flavours. NRQCD results are circles, SESAM results, a star.
The solid
line represents the experimental result. }
\end{figure}

The NRQCD collaboration now has results at $\beta$ values
of 5.7, 6.0 and 6.2 in the quenched approximation (see table 1)
{}~\cite{scaling}.
 Figure 3 shows the scaling of
the ratio of the 2S-1S to 1P-1S splittings. There is no sign
of significant scaling violations and the result clearly disagrees
with experiment. These results use the action with $\delta H_1$
described above in which
the leading $\cal{O}$$(M_Qv^2)$ terms are corrected for their
lowest order discretisation errors. The remaining leading discretisation
error is then expected to be the difference of terms of
$\cal{O}$$(a^4M_Q^5v^6)$ between $s$ and $p$ states.
At $\beta$ = 5.7 the $\cal{O}$$(a^2)$ errors
arising from the gluon field configurations generated with the unimproved
Wilson plaquette action become significant. They can be corrected
for perturbatively ~\cite{alpha} and this has been done in Figure 3.
It amounts to a 5\% reduction in the 1P-1S splittings at $\beta$ = 5.7,
less than one $\sigma$ in the ratio.

Figure 4 shows the ratio of the 1P-1S splitting in the $\Upsilon$
system to the UKQCD values of the $\rho$ mass at the
 same three values of
$\beta$. The UKQCD results shown are those for the tadpole-improved
clover ~\cite{kenway} light fermions. Good scaling is seen and
very clear disagreement with experiment in the quenched approximation.
For comparison I also show the ratio using the GF11 unimproved light
hadron results ~\cite{GF11} - clear violations of scaling are seen.

Having demonstrated scaling of the spin-independent spectrum at
$n_f$ = 0 (and therefore presumably at other values of $n_f$
also), we can now study $n_f$ dependence of the results. Figure 5
shows again the 2S-1S/1P-1S ratio plotted as a function of $n_f$.
The results at $n_f$ = 2 are in much closer agreement with experiment
than at $n_f$ = 0. The two results using different ensembles at
$n_f$ = 2 with different types of dynamical quarks are in good
agreement with each other.

\begin{figure}
\centerline{\ewxy{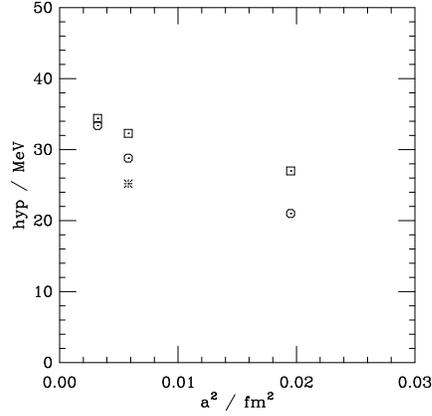}{75mm}}
\vspace{0.2in}
\caption{ The $1^3S_1 - 1^1S_0$ splitting
in MeV with the scale set by
the $2^3S_1 - 1^3S_1$ splitting
in $b\overline{b}$ as a function of the squared lattice spacing in ${\rm fm}^2$
at
3 different values of the lattice spacing in the quenched
approximation. NRQCD collaboration;
the open circles use $u_{0P}$ as in the simulation, the open
squares are rescaled to give the results using $u_{0L}$.
The star gives UKQCD results with a higher order action and
$u_{0P}$.}
\end{figure}

Since important momentum scales in $\Upsilon$ $s$ and $p$ states
are around 1 GeV we might expect these splittings to `see' 3 flavours
of dynamical quarks. Extrapolating linearly through the NRQCD results
at $n_f$ = 0 and 2 does cross the experimental line at $n_f$ = 3.
Improved statistics at $n_f$ = 2 would be useful to show
definitively that this point gives incorrect answers.

The numbers plotted at $n_f$ = 2 are for the lightest dynamical
quark mass in both the NRQCD and SESAM cases. In principle the results
should be extrapolated (possibly linearly ~\cite{grinstein}) in the
light quark mass to the point
\begin{equation}
m_{dyn} = \frac {m_u + m_d + m_s} {3} \approx \frac {m_s} {3} .
\end{equation}
However, no significant $m_{dyn}$ dependence has been seen in either
dynamical ensemble and so this has not been attempted.

\begin{figure}
\centerline{\ewxy{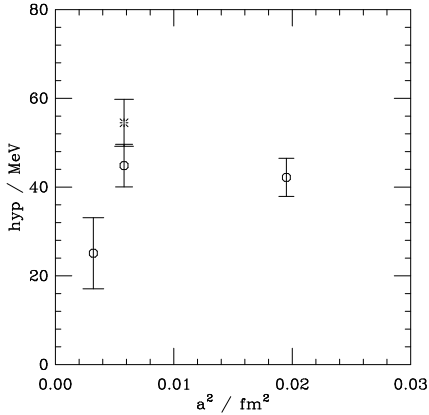}{75mm}}
\vspace{0.2in}
\caption{ The $1^3S_1 - 1^1S_0$ splitting in MeV
with the scale set by
the $1^3P_2 - 1^3P_0$ splitting
in $b\overline{b}$ as a function of the squared
lattice spacing in ${\rm fm}^2$ at
3 different values of the lattice spacing in the quenched
approximation. Open circles, NRQCD collaboration; star UKQCD
collaboration with a higher order action.}
\end{figure}

{\bf Spin-dependent spectrum} -  Figure 2 compares the fine structure
for the sets of results in Table 1, and includes details of the action
used by each group.  It is immediately obvious that
the fine structure is much more sensitive to changes in the quark action.
This is not surprising as the fine structure appears for the first
time at an order one less in the relativistic expansion than the
spin-independent spectrum. $\delta H_1$ used by the NRQCD collaboration
contains only the leading order spin-dependent terms and so we
expect changes at the 10\% level ($v^2 \approx 0.1$) on using $\delta H_2$.
It is also clear that unquenching has a big effect and will be necessary to
get the right answers.

Comparable results with and without $\delta H_2$ are the $\bowtie$
from UKQCD ~\cite{manke} and the $\circ$ from NRQCD, both on quenched
configurations
at $\beta$ = 6.0. A significant (15$\sigma$) effect is seen in the
hyperfine splitting and the shift corresponds to a 10\%
effect. The shifts in the $p$ fine structure may turn out to be
somewhat larger (possibly as much as 20\%) but currently they are not
statistically significant. A comparison of the two collaborations
results with only $\delta H_1$ shows agreement between the two different
methods of smearing.

The $\bowtie$ points include both relativistic corrections (which are
physical) and discretisation corrections (which are not). The scaling
violations in the quenched NRQCD results can attempt to untangle these
effects. Figure 6 shows a scaling plot of the hyperfine
splitting divided by the 2S-1S splitting as a function of $a^2$ in
${\rm fm}^2$. We expect violations of scaling at this order in the
NRQCD results from
errors in the $B$ field in the $\sigma \cdot {\bf B}$ term which gives
rise to this splitting. Figure 6 shows clear scaling violations
with a slope $\mu$ given by $\approx$ 900 MeV, when we write ~\cite{john}:
\begin{equation}
\left( \frac {hyp} {2S-1S} \right) = \left( \frac {hyp} {2S-1S}
\right)_{a=0} \{ 1 - (\mu a)^2 \} .
\end{equation}

It is not surprising to find scaling violations of this size since the
hyperfine splitting responds to much shorter distances than
spin-independent splittings and is given by a single term
in $\delta H$ without the cancellation that occurs for
derivative type discretisation terms in spin-independent splittings.

The hyperfine splitting is also sensitive to the value of
the bare quark mass (like $1/M_Qa$) and errors in how well
this is fixed will affect Figure 6.

In addition there is sensitivity to the value of $u_0$, because the
$B$ field contains 4 links. We expect the hyperfine splitting
to vary as $1/u^6_0$, and this was borne out by work in ref.~\cite{ups}
where results with $u_0$ and without ($u_0$ = 1) were compared.
NRQCD and UKQCD results both use $u_0$ from the fourth root of
the plaquette, denoted $u_{0P}$ in Figure 2. Recent work by Trottier
{}~\cite{trottier}
has suggested that  $u_0$ from the Landau gauge link, $u_{0L}$, might provide a
better estimate of the radiative corrections to $c_4$. This might
also be true for other $c_i$ if $u_{0L}$ captures tadpole
effects more accurately. This value of $u_{0L}$ is therefore
used by the SESAM collaboration and has the effect of increasing
the hyperfine splitting over the value that would be obtained with
$u_{0P}$.

Figure 6 gives NRQCD values for the ratio of hyperfine to 2S-1S splittings
that might be expected using $u_{0L}$ by rescaling the results
obtained with $u_{0P}$ by the sixth power of the ratio of $u_0$
values. The $u_{0L}$ results are somewhat flatter and this might
indicate that some of the previous scaling violations arose
from radiative corrections to $c_4$. Perturbative or non-perturbative
calculations of various $c_i$ will be needed to answer the question of
which $u_0$ is better (if there is a single answer) and allow us
to include radiative corrections in the leading coefficients in a
consistent next-to-leading-order calculation of fine structure.

\begin{figure}
\centerline{\ewxy{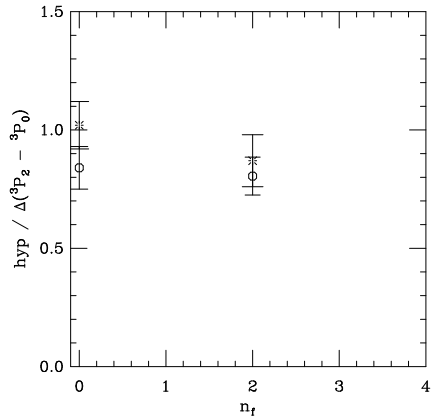}{75mm}}
\vspace{0.2in}
\caption{ The ratio of the $1^3S_1 - 1^1S_0$ splitting
to the $1^3P_2 - 1^3P_0$ splitting
in $b\overline{b}$ as a function of $n_f$.
Open circles, NRQCD collaboration; star, UKQCD collaboration
at $n_f$ =0, SESAM collaboration at $n_f$ = 2. See text. }
\end{figure}

In $n_f$ extrapolations of fine structure splittings it might not
be true that $n_f$ = 3 is the relevant physical point. $n_f$ = 4
is somewhat more likely given the short distance nature of these
quantities. In that case the extrapolations in $n_f$ must
be done in terms of other quantities for which $n_f$ = 4 is
also the physical point, and {\it not} the 1P-1S or 2S-1S
splittings. Indeed, there will be no value of degenerate
dynamical $n_f$ for which we could get the right answer for
the ratio of fine structure to spin-independent splittings.
Instead, for example,  we would have to extrapolate the ratio of
the hyperfine splitting to one of the fine structure splittings.
For this ratio better scaling is seen in the NRQCD
results, as shown in Figure 7, although the results at
$\beta$ = 6.2 are still rather uncertain. So an extrapolation in $n_f$
could then be done.

The results at different $n_f$ are shown in Figure 8, where NRQCD results
are compared at $n_f$ = 0 and 2, and UKQCD results at $n_f$ = 0
are compared to SESAM results at $n_f$ = 2. The SESAM
results are not strictly comparable with UKQCD since they use
$u_{0L}$; if $u_{0P}$ had been used their result for the ratio
plotted would be smaller by about 5\%.
 Unfortunately the fine structure
splittings are not very accurate and the resulting error in
extrapolating the hyperfine splitting in this way is very large.
At $n_f$ = 4, the hyperfine splitting would be 40(10) MeV.

A comparison of the $p$ fine structure to experiment from
Figure 2 shows interesting features, albeit with
large errors. The overall scale of the splittings given
by $^3P_2 - ^3P_0$
is obviously too small in the quenched approximation
(and is reduced further when scaling violations are removed as for the
hyperfine splitting). The unquenched results are much closer
to experiment.

The ratio of splittings, $(^3P_2 - ^3P_1)/(^3P_1 - ^3P_0)$,
is larger than experiment for the NRQCD results, 1.1(4) versus
0.66(2). That this is probably a discretisation
error is borne out by the apparent improvement in this ratio
in the UKQCD results. The ratio viewed in a potential model
picture has differences of long and short range effects in it
and it would be unlikely to be correct in the quenched approximation,
but is hard to determine accurately. The unquenched results in
Figure 2 look encouragingly in agreement with experiment but
a full study of scaling and $n_f$ extrapolations must be done.

Obviously more work is required on the fine structure of
the $b\overline{b}$ system and I believe that NRQCD will
provide the most accurate results in this area. It will also
be important to understand which quantities should be
extrapolated to which values of $n_f$ or, failing this,
to make configurations with `real world' dynamical quark content.
Only then will we be able to get the correct answer for
all the splittings without $n_f$ extrapolations.

\begin{figure}
\centerline{\epsfig{file=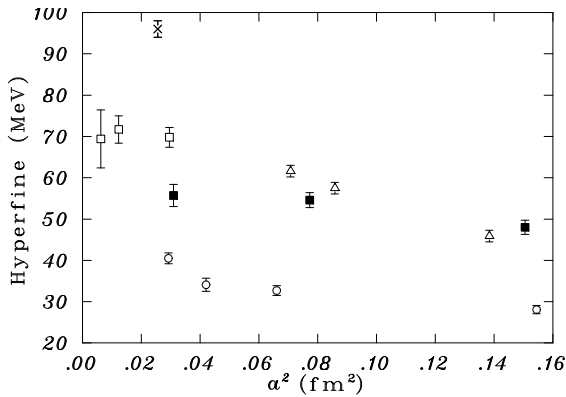,height=75mm,bbllx=115pt,bblly=133pt,
bburx=476pt,bbury=660pt,angle=90,clip=}}
\vspace{0.2in}
\caption[jgkfgj]{
The hyperfine splitting for charmonium versus $a^2$ taken
from ref. \protect ~\cite{trottier}. The next-to-leading order NRQCD
results are given with $u_{0P}$, open circles, and with $u_{0L}$, filled
boxes. The leading order NRQCD result \protect ~\cite{charm} is marked with a
cross.
The relativistic heavy Wilson approach \protect ~\cite{fnal} gives the open
boxes. }
\end{figure}

\section{Results for Charmonium spectroscopy}

The charmonium sector is a difficult one to simulate on the
lattice, because it is neither very relativistic or
very non-relativistic. For NRQCD to remain above $M_ca$ of around 0.6
requires $\beta < 5.85$ ~\cite{charm}. Each order of relativistic
correction is only 30\% smaller than the previous order since
$v^2 \approx 0.3$.
For the standard relativistic heavy Wilson approach ~\cite{fnal}
the opposite requirement is true, $\beta \ge 6.0$,
because of problems with fixing the meson mass as a shift
from the energy at zero momentum ~\cite{sara}.

There has been a long standing issue ~\cite{me_lat93} that results for
charmonium hyperfine splittings disagreed between
these two methods, with the NRQCD results giving a
hyperfine splitting of around 100 MeV and the relativistic
method around 70 MeV (experiment 116 MeV).

New work by Trottier in NRQCD ~\cite{trottier} using the relativistic and
discretisation corrections of $\delta H_2$ has shown that
indeed the relativistic corrections from higher order terms
in NRQCD are large. He finds the hyperfine splitting becomes
smaller
between the $\delta H_1$ results and the $\delta H_2$ results
by a factor of 60\%. This is certainly of $\cal{O}$$(30\%)$
so not in principle surprising but nevertheless disappointing
to proponents of NRQCD.

A shift of the same size in spin-independent splittings would be
a 15\% effect (certainly of the same order as the
naive expectation of 10\%) since they are lower order in the non-relativistic
expansion. This means a systematic error in $a^{-1}$ determinations
from NRQCD charmonium at this level
if next-to-next-to-leading order spin-independent
terms are not included. This is significantly less than the statistical
error that can be achieved, but might nevertheless be acceptable in
some applications.

It seems likely that future progress in the charmonium sector
will use the heavy Wilson approach, possibly corrected for
$p^4$ terms, as has been suggested ~\cite{kronfeld}.

\section{ Results for $B_c$ spectroscopy}

Recent experimental evidence for the $B_c$ particle encourages
lattice predictions for the spectrum of $b\overline{c}$ bound
states. Early work ~\cite{bc} has used quenched configurations
and NRQCD for both the $b$ and the $c$ quarks.

More recent results are collected in Figure 10.
These include NRQCD collaboration results analysed by Martin Gorbahn
on dynamical configurations
from the MILC collaboration ($\beta$ = 5.415, $n_f$ = 2, KS, $ma$
=0.0125) and UKQCD results from Hugh Shanahan using NRQCD $b$ and relativistic
(tadpole-improved clover) $c$ quarks on $\beta$ = 6.2 quenched
configurations. For comparison the older results ~\cite{bc} are
shown and results from a potential model analysis ~\cite{equigg}.
There is no clear discrepancy with the potential model results as
yet, despite the fact that the $c$ quark is more relativistic
in $B_c$ than in charmonium.

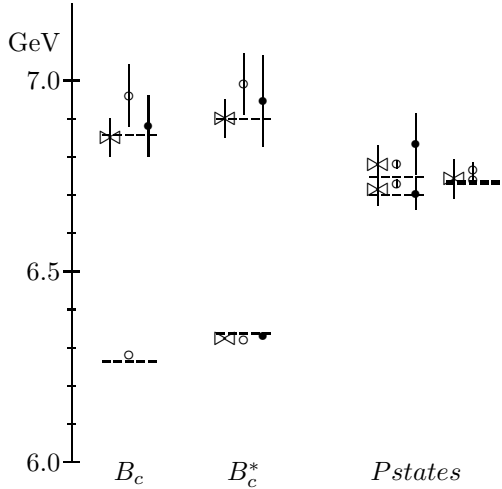
\begin{figure}
\begin{center}
\setlength{\unitlength}{.02in}
\begin{picture}(130,120)(10,580)
\put(15,600){\line(0,1){120}}
\multiput(13,600)(0,50){3}{\line(1,0){4}}
\multiput(14,600)(0,10){11}{\line(1,0){2}}
\put(12,600){\makebox(0,0)[r]{6.0}}
\put(12,650){\makebox(0,0)[r]{6.5}}
\put(12,700){\makebox(0,0)[r]{7.0}}
\put(12,710){\makebox(0,0)[r]{GeV}}

     \put(30,600){\makebox(0,0)[t]{$B_c$}}
     \put(30,628){\circle{2}}
     \multiput(23,626.4)(3,0){5}{\line(1,0){2}}

     \put(30,696){\circle{2}}
     \put(30,696){\line(0,1){8}}
     \put(30,696){\line(0,-1){8}}
     \multiput(23,685.6)(3,0){5}{\line(1,0){2}}
     \put(25, 685){\makebox(0,0){$\bowtie$}}
     \put(25,685){\line(0,1){5}}
     \put(25,685){\line(0,-1){5}}
     \put(35,688){\circle*{2}}
     \put(35,688){\line(0,1){8}}
     \put(35,688){\line(0,-1){8}}

     \put(60,600){\makebox(0,0)[t]{$B^{*}_c$}}
     \put(60,632){\circle{2}}
     \multiput(53,633.7)(3,0){5}{\line(1,0){2}}
     \put(55, 632.3){\makebox(0,0){$\bowtie$}}
     \put(65,633){\circle*{2}}

     \put(60,699){\circle{2}}
     \put(60,699){\line(0,1){8}}
     \put(60,699){\line(0,-1){8}}
     \multiput(53,689.9)(3,0){5}{\line(1,0){2}}
     \put(55, 690){\makebox(0,0){$\bowtie$}}
     \put(55,690){\line(0,1){5}}
     \put(55,690){\line(0,-1){5}}
     \put(65,694.6){\circle*{2}}
     \put(65,694.6){\line(0,1){12}}
     \put(65,694.6){\line(0,-1){12}}


     \put(105,600){\makebox(0,0)[t]{$P states$}}
     \put(100,673){\circle{2}}
     \put(100,673){\line(0,1){1}}
     \put(100,673){\line(0,-1){1}}
     \multiput(93,674.7)(3,0){5}{\line(1,0){2}}
     \put(95, 671.3){\makebox(0,0){$\bowtie$}}
     \put(95,671.3){\line(0,1){4}}
     \put(95,671.3){\line(0,-1){4}}
     \put(105,670.3){\circle*{2}}
     \put(105,670.3){\line(0,1){4}}
     \put(105,670.3){\line(0,-1){4}}

     \put(100,678){\circle{2}}
     \put(100,678){\line(0,1){1}}
     \put(100,678){\line(0,-1){1}}
     \multiput(93,670.0)(3,0){5}{\line(1,0){2}}
     \put(95, 677.8){\makebox(0,0){$\bowtie$}}
     \put(95,677.8){\line(0,1){5}}
     \put(95,677.8){\line(0,-1){5}}
     \put(105,683.3){\circle*{2}}
     \put(105,683.3){\line(0,1){8}}
     \put(105,683.3){\line(0,-1){8}}

     \put(120,674){\circle{2}}
     \put(120,674){\line(0,1){1}}
     \put(120,674){\line(0,-1){1}}
     \put(120,676.5){\circle{2}}
     \put(120,676.5){\line(0,1){2}}
     \put(120,676.5){\line(0,-1){2}}
     \multiput(113,673.0)(3,0){5}{\line(1,0){2}}
     \multiput(113,673.6)(3,0){5}{\line(1,0){2}}
     \put(115, 674.2){\makebox(0,0){$\bowtie$}}
     \put(115,674.2){\line(0,1){5}}
     \put(115,674.2){\line(0,-1){5}}


\end{picture}
\end{center}
\caption{Lattice results for the spectrum of the
$B_c$ system. Open circles are results using NRQCD for
both $b$ and $c$ quarks on quenched configurations
at $\beta$ = 5.7. Filled circles use unquenched
MILC configurations.
$\bowtie$ are results using NRQCD $b$ quarks and
heavy Wilson $c$ quarks on quenched configurations
at $\beta$ = 6.2.
Error bars represent statistical uncertainties only.
Dashed lines show results from a recent
potential model calculation \protect \cite{equigg}}
\label{fig:bc}
\end{figure}

One of the problems with a mixed system like the $B_c$ is an
ambiguity in what to take for the quark masses. This is
quite significant in the quenched approximation because the
value of $a^{-1}$ depends on whether it is fixed from the
$\Upsilon$ or $\Psi$ system and the values of the bare
quark masses have also been fixed separately within these systems.
The splittings in Figure 10 will change somewhat if the
quark masses were altered.
To perform
consistent extrapolations in $n_f$ it will be necessary to
use one particular splitting in, say, the $\Upsilon$ system to
fix $a^{-1}$ and then fix $m_c$ using this $a^{-1}$ in the $\Psi$
system. This has not been done as yet.

One interesting feature of the $B_c$ system is its similarity
to heavy-light systems, allowing a test of some of the
techniques that will be useful for the spectrum there.
In particular the spin 1 $p$ states will mix because of
a lack of charge conjugation. It was possible for the
quenched NRQCD results ~\cite{bc} to diagonalise the mixing
matrix and pick out the physical $1^{+}$ and $1^{+'}$ states.
This will also need to be done in the $B$ sector, but has not
been possible as yet ~\cite{arifa}.

\vspace{4mm}

{\bf Acknowledgements}\\
I would like to thank the organisers for a very useful meeting.
I am grateful to my collaborators, Peter Lepage, Paul
McCallum, Junko Shigemitsu and John Sloan, in the NRQCD collaboration
for many useful discussions.
I am also grateful to Achim Spitz and Henning Hoeber (SESAM), Thomas Manke
and Hugh Shanahan (UKQCD), Howard Trottier and Martin Gorbahn for discussions
of their results.

\end{document}